\documentclass[12pt,a4,fleqn]{article}
\usepackage{srcltx}
 \usepackage{fancyhdr}
\usepackage[cp1251]{inputenc}
\usepackage[english]{babel}
\usepackage[T2A]{fontenc}
\usepackage{amsmath,amssymb,epsfig,amsfonts,amscd}
\usepackage{changebar}
\usepackage{graphicx}
\usepackage{color}

\begin{document}
\begin{center}
\large {\bf Deformed gas of $p,q$-bosons: virial expansion and virial coefficients}
\end{center}

\begin{center}
A.M. Gavrilik\footnote{omgavr@bitp.kiev.ua}, A.P. Rebesh

\vspace{1mm} {\it Bogolyubov Institute for Theoretical Physics,
               Kiev, Ukraine}

\end{center}







\begin{abstract}
In the study of many-particle systems both the interaction of
particles can be essential and such feature as their internal
(composite) structure. To describe these aspects, the theory of
deformed oscillators is very efficient. Viewing the particles as
$p,q$-deformed bosons, in the corresponding $p,q$-Bose gas model we
obtain in explicit form virial expansion along with the 2nd to 5th
virial coefficients. The obtained virial coefficients depend on the
deformation parameters $p,q$ in the form symmetric under
$p\leftrightarrow q$, and at $p\rightarrow1$, $q\rightarrow1$ turn
into those known for usual bosons. Besides real parameters, we
analyze the case of complex mutually conjugate $p$ and $q$ and find
interesting implications. Also, the critical temperature is derived
(for the $p,q$-Bose gas) and compared with the $T_c$ of standard
case of bosons condensation. Similar results are presented for the
deformed Bose gas model of the Tamm-Dancoff type.
\end{abstract}

\hspace{7mm} {\bf Keywords:} deformed Bose gas model, virial
expansion, virial coefficients,

\hspace{31mm} critical temperature

\section{\label{sec:1} Introduction}
Diverse models of deformed Bose gas, elaborated on the base of a set
of some deformed oscillators or deformed bosons, appeared in the
literature from early nineties, see e.g. Refs.
\cite{Martin91}-\cite{GangMo92}. Some analysis of a number of works
on this topic has been done in Ref. \cite{Vokos1994}. Since then,
the statistical mechanics of a gas of $q$- or $p,q$-bosons was
extensively explored (and further extended) as witnessed by
\cite{Avancini94}-\cite{GavrReb2011} and many others.

With the aim of treating possible intermediate statistics behavior of
physical system, one can use the method based on a deformation of quantum
algebra of creation, annihilation and number operators.
Deformed oscillator algebra is a generalization of the harmonic oscillator or Heisenberg
algebra.
   Besides intermediate statistics, there exist other convincing reasons of why
   one should use some model of deformed oscillator instead of the
usual quantum harmonic oscillator. The models of deformed oscillators are
efficient in description of
interacting particles system, or if the finite proper volume of particles, or their
composite nature (internal structure) are to be taken into
account.
   The theory of $q$-oscillators
is related to the theory of quantum groups as originally shown in
Ref. \cite{Biedenharn1989} and Ref. \cite{Macfarlane1989}. As a
direct generalization of $q$-oscillators, in Ref. \cite{chakjag1991}
and Ref. \cite{ArikZ} the two-parameter family of $p,q$-oscillators
has been introduced.

The two-parameter $p,q$-deformed analog of Bose gas model can be
developed, see \cite{AdG04}, \cite{GavrSig06}, using the set of
$p,q$-deformed oscillators. Some further aspects of that model were
studied in the works \cite{Algin2010}, \cite{Algin2008}. In our
present work we explore the thermodynamics of (some version of)
$p,q$-Bose gas model. In order to obtain certain thermodynamical
quantities of the deformed Bose gas we use the two-parametric
generalization of $q$-calculus realized by replacing the ordinary
derivative with the $p,q$-analog of Jackson derivative (the latter
is known as $q$-deformed analog of usual derivative).

For the $p,q$-deformed Bose gas we investigate two regimes: high
temperatures and small densities ($\lambda^2/{\it v}\!\ll\!1$), or
low temperatures and large densities ($\lambda^2/{\it v}\!\gg1\!$).
In the first case we obtain explicitly the virial expansion of the
equation of state and find new virial coefficients in addition to
few known ones (mainly in more special cases of one-parameter
deformed Bose gas models \cite{Monteiro93}, \cite{Scarfone}). The
virial coefficients reflect effective inter-particle interactions
and in the considered case depend explicitly on the deformation
parameters $p$ and $q$. From our formulas, the ordinary boson gas
results can be recovered in the corresponding limit $p\rightarrow1$,
$q\rightarrow1$. On the other hand, setting $p$ and $q$ as complex
conjugates: $p\!=\!\overline{q}\!=\!re^{i\theta}$, we gain yet
another presentation of the virial coefficients. In that case we
examine the behavior of the difference between deformed coefficients
and their ordinary non-deformed prototypes (corresponding to {\it
non-deformed} Bose gas) as function of the new parameter $\theta$,
at fixed values of the parameter $r$ (the modulus). The situation
for which Bose-Einstein condensation does occur in the considered
two-parameter generalized boson gas is studied when $\lambda^2/{\it
v}\gg1$, and in this case the critical temperature $T_c^{(p,q)}$ of
the condensation of deformed bosons is obtained. Comparing this with
the $T_c$ of usual Bose gas we deduce that the critical temperature
of the $p,q$-Bose gas is larger than the critical temperature of
usual bosons, for the considered range of parameters. The dependence
of the ratio $T_c^{(p,q)}\!/\,T_c$ on the deformation parameters
$p,q$ is explicitly studied, visualized by the corresponding figures
and the distinction from the results in Ref.
 \cite{Algin2008} indicated. We also consider similar results for
the $q$-deformed Bose gas model of Tamm-Dancoff (TD) type, which
constitutes a distinguished special case.

\section{\label{sec:2} Deformed oscillator models}
In the theory of deformed oscillators it is convenient to use the
concept \cite{Meljanac} of deformation structure function
$\varphi(N)$ which determines the particular deformed oscillator
model described by its respective oscillator algebra. With
$\varphi(N)\equiv a^{\dagger}a$, the generating elements $a$,
$a^{\dagger}$, $N$ of the algebra obey the relations
\begin{equation}\label{1}
[N,a^{\dagger}]\!=\!a^{\dagger}, \hspace{5mm}
[N,a]\!=\!-a,
\end{equation}
\begin{equation}\label{2}
aa^{\dagger}\!-\!a^{\dagger}a\!=\!\varphi(N\!+\!1)\!-
\!\varphi(N), \hspace{8mm} aa^{\dagger}=\varphi(N\!+\!1).
\end{equation}
In deformed Fock space the ground state vector obeys usual relations
\begin{equation}\label{3}
a|0\rangle\!=\!0, \hspace{5mm} N|0\rangle\!=\!0,
    \hspace{5mm} \langle0|0\rangle\!=\!1,
\end{equation}
and for the $n$-particle excited state in  this
$\varphi(n)$-extended Fock space we have
\begin{equation}\label{4}
N|n\rangle\!=\!n|n\rangle, \hspace{5mm} \varphi(N)|n\rangle\!\!=\!
\varphi(n)|n\rangle,
\end{equation}
   \begin{equation}\label{5}
 |n\rangle\!=\!\frac{(a^{\dagger})^n}{\sqrt{\varphi(n)!}}|0\rangle,
\hspace{3mm} \varphi(n)!\!=\!\varphi(n)\varphi(n\!-\!1)...\varphi(1),
\hspace{3mm} \varphi(0)!\!=\!1.
\end{equation}
Operators $a^{\dagger}$, $a$ act on the state $|n\rangle$ according to
\begin{equation}\label{6}
a^{\dagger}|n\rangle\!=\!\sqrt{\varphi(n\!+\!1)}|n\!+\!1\rangle,
\hspace{5mm} a|n\rangle\!=\!\sqrt{\varphi(n)}|n\!-\!1\rangle .
\end{equation}
In this paper we deal with the system of so-called $p,q$-bosons ($p,q$-deformed oscillators)
for which the deformation structure function $\varphi(n)$ in (\ref{2})-(\ref{6}) is
\begin{equation}\label{7}
\varphi_{p,q}(n)=\frac{p^n-q^n}{p-q}\equiv[n]_{p,q},
\end{equation}
with $[n]_{p,q}$ being the $p,q$-number corresponding to a number $n$.
Let us note that from the two-parameter family one infers a plenty
of different one-parameter ($q$-deformed) models by imposing different relations
$p=f(q)$, see \cite{Plethora} for more details.

\section{\label{sec:3} Thermodynamics of the gas of deformed bosons}
In this Section we deal with a particular variant of the
$p,q$-deformed Bose gas model. In the grand canonical ensemble, the
Hamiltonian of the system is taken as
\begin{equation}\label{8}
H\!=\!\displaystyle\sum_i(\varepsilon_i\!-\!\mu)N_i
\end{equation}
where $\varepsilon_i$ is the (kinetic) energy of particle in the state labeled by $i$,
$N_i$ is the boson number operator relative to
$\varepsilon_i$ and $\mu$ is the chemical potential.
One-particle non-relativistic energies are given as $\varepsilon_i=p_i^2/2m$
where $p\equiv |\overrightarrow{p}|$.
Thermal averages can be calculated using the usual formulas
of quantum statistical mechanics.
 Therefore, the thermal average
of an operator $A$ is given by the standard formula
\begin{equation}\label{9}
\langle \widehat{A} \rangle=Tr(\rho\widehat{A}), \hspace{10mm} \rho=\frac{e^{-\beta\widehat{H}}}{Z}
\end{equation}
where $\rho$ is the equilibrium statistical operator, $Z\!=\!Z(z)$ is
the partition function,
\begin{equation}\label{10}
Z\!=\!Tr(e^{-\beta \widehat{H}}).
\end{equation}
Besides, $\beta=1/kT$, $T$ is the temperature;
the Boltzmann constant $k$ is put $k=1$.

To develop thermodynamics of the gas of deformed bosons we start with the
logarithm of grand partition function
\begin{equation}\label{11}
\ln Z=-\displaystyle\sum_i\ln(1\!-\!ze^{\!-\beta\varepsilon_i})
\end{equation}
where $z\!=\!e^{\beta\mu}$ is the fugacity.
 For obtaining
thermodynamic functions of $q$-deformed Bose gas the usual
derivative is replaced by the Jackson or $q$-derivative
\cite{Jackson}
\begin{equation}\label{12}
\frac{d^{(q)}}{dx}f(x)\!=\!\frac{f(qx)\!-
\!f(x)}{qx\!-\!x}{\equiv{\mathcal{D}_x^{(q)}}}f(x).
\end{equation}
Note that in the limit $q\to 1$ the usual derivative is recovered
from the $q$-derivative.

For the algebra given by (\ref{1}), (\ref{2}) and $\varphi(n)$ from
(\ref{7}) we use the two-parameter $p,q$-generalization of Jackson
derivative\footnote{Let us note that in Refs. \cite{Algin2010},
\cite{Algin2008} yet another two-parameter analog
$\widetilde{D}^{(p,q)}_x$ of the Jackson derivative is exploited
which is related with $D^{(p,q)}_x$ in (\ref{13}) as follows:
$\widetilde{D}^{(p,q)}_x\!=\!\frac{p-q}{\ln p-\ln q}D^{(p,q)}_x$.}
being a direct $p,q$-extension of (\ref{12}), i.e.
\begin{equation}\label{13}
{\mathcal{D}_x^{(p,q)}}f(x)=\frac{f(px)\!-\!f(qx)}{px\!-\!qx},
\qquad \quad
{\mathcal{D}_x^{(p,q)}}x^n\!=\!\frac{p^nx^n-q^nx^n}{px-qx}\!=\!
[n]_{p,q}x^{n-1}.
\end{equation}
When $p=1$, the latter $p,q$-extension reduces to the Jackson
derivative (\ref{12}).

Then, the total number of particles in the $p,q$-deformed theory is determined
as
\begin{equation}\label{14}
N\!=\!z{\mathcal{D}_z^{(p,q)}}\ln Z\!=\!\frac{1}{p-q}
\Bigr(\!\!\displaystyle\sum_i\ln(1\!-\!qze^{-\beta\varepsilon_i})\!-\!
\displaystyle\sum_i\ln(1\!-\!pze^{-\beta\varepsilon_i})\!\!\Bigl).
\end{equation}

For large volume $V\!\rightarrow\!\infty$ and large number of particles $N$,
we replace summation over level $i$ by integration over the 3-momentum $k$.
   Note that the two sums in (\ref{14})
diverge at $z=1$ and $p\to 1$, $q\to 1$ when $\varepsilon_i\!=\!0$.
That is, these $\varepsilon_i\!=\!0$ terms can be as large as all
the rest of terms in the sums. For that reason the
$\varepsilon_i\!=\!0$ term is isolated, while the remaining two sums
are replaced by integrals:
\begin{equation}\label{15}
\displaystyle\sum_i \ \rightarrow \ \frac{V}{(2\pi\hbar)^3}\int d^3k.
\end{equation}
Going over in $\overrightarrow{k}$-space to spherical coordinates
$\varphi$, $\theta$,
$k=(\overrightarrow{k}\cdot\overrightarrow{k})^{1/2}$, the
integration is performed with the result given through
polylogarithms. Since the latter can be presented in the form of
series, we derive the result

\begin{equation}\label{16}
N\!=\!\frac{V}{\lambda^3}\Bigr[\frac{1}{p-q}
\biggr(\!\sum_{r=1}^{\infty}
\frac{(zp)^r}{r^{5/2}}\,-\!\sum_{r=1}^{\infty}
\frac{(zq)^r}{r^{5/2}}\biggl)\Bigl]+\frac{1}{p-q}\,ln\Bigr(\frac{1\!-
\!qz}{1\!-\!pz}\Bigl)=\frac{V}{\lambda^3}\sum^{\infty}_{r=1}
\frac{[r]_{p,q}}{r^{5/2}}z^r+\,n_0 \ \
\end{equation}
where $n_0=\frac{1}{p-q}\ln\bigl(\frac{1-qz}{1-pz}\bigr)$ and $\lambda\!=\!\sqrt{\frac{2\pi\hbar^2}{mT}}$ is the thermal wavelength.
Now rewrite Eq. (\ref{16}) as
\begin{equation}\label{17}
\frac{1}{\it{v}}\!=\!
\frac{1}{\lambda^3}\,{\widetilde{g}_{3/2}}(z;p,q)+\frac{n_0}{V}
\end{equation}
where $\it{v}\!=\!V/N$, and the general formula for the
$(p,q)$-extended function ${\widetilde{g}_{n}}$ reads:
\begin{equation}\label{18}
{\widetilde{g}_{n}}(z;p,q)\!=\!\frac{1}{p-q}\biggr(\sum_{r=1}^{\infty}
\frac{(zp)^r}{r^{n+1}}-\sum_{r=1}^{\infty}
\frac{(zq)^r}{r^{n+1}}\biggl)=\sum^{\infty}_{r=1}\frac{[r]_{p,q}z^r}{r^{n+1}}.
\end{equation}
This function is nothing but $p,q$-deformed generalization of the $g_n(z)$
function known say from Ref. \cite{Huang}:
\begin{equation}\label{19}
g_n(z)\equiv\sum^{\infty}_{l=1}\frac{z^l}{l^n}.
\end{equation}
The function (\ref{18}) is real for real $p$ and $q$. We assume
$0<p\leq1$, $0<q\leq1$. In what follows we will use
$\widetilde{g}_{3/2}$ and $\widetilde{g}_{5/2}$. Namely, we will
exploit (for small $z$) their power series expansions, say up to
$z^5$ order.

Then the thermodynamic relation $PV\!/\,T=\ln Z$ can be rewritten as
\begin{equation}\label{20}
\frac{P}{T}\!=\!
\frac{1}{\lambda^3}\,{\widetilde{g}_{5/2}}(z;p,q)+\frac{1}{V}\ln(1-z).
\end{equation}

\subsection{Virial expansion of the equation of state}
In the case of high temperatures and small densities
($\lambda^3/{\it v}\ll1$), the average distance ${\it v}^{1/3}$
between particles is much larger then the thermal wavelength
$\lambda$. In this case the quantum effects, the $n_0$, and the
second term in (\ref{20}) can be viewed as negligibly small. Then,
if the temperature of the gas of $p,q$-deformed bosons is high, so
that $T>T_c^{(p,q)}$ (see Sect. \ref{sec:4} below for
$T_c^{(p,q)}$), from (\ref{17}) and (\ref{20}) we find
\begin{equation}\label{21}
\frac{1}{\it{v}}\!=
\!\frac{1}{\lambda^3}\,{\widetilde{g}_{3/2}}(z;p,q), \hspace{10mm}
\frac{P\it{v}}{T}\!=
\!\frac{\it{v}}{\lambda^3}\,{\widetilde{g}_{5/2}(z;p,q)}.
\end{equation}
From the first equation in (\ref{21}) with account of (\ref{18}) we
obtain
\begin{equation}\label{22}
\frac{\lambda^3}{\it{v}}\!=\!z+\frac{[2]_{p,q}}{2^{5/2}}z^2+
\frac{[3]_{p,q}}{3^{5/2}}z^3+\frac{[4]_{p,q}}{4^{5/2}}z^4+\frac{[5]_{p,q}}{5^{5/2}}z^5+... .
\end{equation}
Inverting the latter equation yields
$$
z\!=\!\frac{\lambda^3}{\it
v}\,-\frac{[2]_{p,q}}{2^{5/2}}\biggl(\!\frac{\lambda^3}{\it
v}\!\biggr)^2\!\!+
\biggl(\frac{[2]^2_{p,q}}{2^4}-\frac{[3]_{p,q}}{3^{5/2}}\!\biggr)\!
\biggl(\frac{\lambda^3}{\it v}\biggr)^3\!\!-
\biggl(\frac{5[2]^3_{p,q}}{2^{15/2}}-\frac{5[2]_{p,q}[3]_{p,q}}{2^{5/2}3^{5/2}}+
\frac{[4]_{p,q}}{2^5}\!\biggr)\!\biggl(\frac{\lambda^3}{\it
v}\biggr)^4\!+
$$
\begin{equation}\label{23}
+\biggl(\!\frac{7[2]^4_{p,q}}{2^{9}}+\frac{7[2]^2_{p,q}[3]_{p,q}}{2^{5}3^{3/2}}-
\frac{[3]^2_{p,q}}{3^4}+
\frac{3[2]_{p,q}[4]_{p,q}}{2^{13/2}}+\frac{[5]_{p,q}}{5^{3/2}}\!\biggr)\!
\biggl(\!\frac{\lambda^3}{\it v}\biggr)^5.
 \end{equation}
From the second equation in (\ref{21}), with $\widetilde{g}_{5/2}(z;p,q)$ from (\ref{18})
taken in the form
expanded in $z$, by the use of (\ref{23}) we derive for
the equation of state of the deformed $p,q$-Bose gas the desired virial expansion
\begin{equation}\label{24}
\frac{P\it{v}}{T}=1+A\biggr(\frac{\lambda^3}{\it{v}}\biggl)+
B\biggr(\frac{\lambda^3}{\it{v}}\biggl)^{\!2}+
\,C\biggr(\frac{\lambda^3}{\it{v}}\biggl)^{\!3}+
D\biggr(\frac{\lambda^3}{\it{v}}\biggl)^{\!4}+...
\end{equation}
where the virial coefficients from 2nd to 5th read:
$$A=-\frac{[2]_{p,q}}{2^{7/2}}, \hspace{7mm}
B=\frac{[2]_{p,q}^2}{2^{5}}-\frac{2[3]_{p,q}}{3^{7/2}}, \hspace{7mm}
C=\frac{[2]_{p,q}[3]_{p,q}}{2^{5/2}3^{3/2}}-\frac{3[4]_{p,q}}{2^{7}}-
\frac{5[2]_{p,q}^3}{2^{17/2}},
$$
$$
D=\frac{7[2]^4_{p,q}}{2^{10}}-\frac{[2]^2_{p,q}[3]_{p,q}}{2^{4}3^{5/2}}+
\frac{[2]_{p,q}[4]_{p,q}}{2^{11/2}}+
\frac{2[3]^2_{p,q}}{3^{5}}-\frac{4[5]_{p,q}}{5^{7/2}}.
$$
Note, in the just obtained expressions for virial coefficients, $p$
and $q$ appear only through $(p,q)$-numbers $[2]_{p,q}$,
$[3]_{p,q}$, etc. Due to that, the exchange $p\leftrightarrow q$
symmetry remains intact. Let us note that the above expressions for
$A$, $B$, $C$, $D$ yield the results for the particular cases of
$q$-bosons. Namely, at $p=1$ these reduce to the virial coefficients
of the AC type $q$-Bose gas, and putting $p=q^{-1}$ yields the BM
type $q$-Bose gas virial expansions and virial coefficients. Another
distinguished TD type case is considered separately in Sec.
\ref{sec:5} below. Also, it should be noted that by putting $p=f(q)$
where $f(q)$ is some fixed function, as considered in Ref.
\cite{Plethora}, we readily obtain the relevant results (virial
coefficients) for the corresponding, to this choice of $f(q)$,
version of $q$-deformed Bose gas model.

Remark that the one-parameter ($q$-deformed) version of the above
$A$, $B$, $C$ was given in \cite{Scarfone}, \cite{Monteiro93}. The
$p,q$-deformed 3-rd, 4-th and 5-th virial coefficients $B$, $C$ and
$D$ are new. Note also that our $A=A(p,q)$ differs from the
respective coefficient in \cite{Algin2010}. In the limiting
no-deformation case $p=q=1$ from the deformed virial coefficients
$A$, $B$, $C$, $D$ one recovers virial coefficients $A_0$, $B_0$,
$C_0$, $D_0$ of the usual Bose gas:
 $$A\stackrel{p=q=1}{\longrightarrow}A_0=-\frac{1}{2^{5/2}}, \hspace{12mm} C\stackrel{p=q=1}{\longrightarrow}C_0=\frac{1}{2^{3/2}3^{1/2}}-\frac{3}{2^{5}}-\frac{5}{2^{11/2}},
$$
$$
B\stackrel{p=q=1}{\longrightarrow}B_0=\frac{1}{8}-\frac{2}{3^{5/2}}, \hspace{7mm}
D\stackrel{p=q=1}{\longrightarrow}D_0=\frac{7}{2^{6}}-\frac{1}{2^2\cdot3^{3/2}}+\frac{1}{2^{5/2}}+\frac{2}{3^{3}}-\frac{4}{5^{5/2}}.
$$
It is seen that $A_0$, $B_0$ and $C_0$ coincide with the well-known
virial coefficients of Bose gas given in textbooks \cite{Kubo65},
\cite{Pathria}, as it should.

\subsection{Complex parameters of deformation and virial coefficients}
Now take the deformation parameters $p$, $q$ as complex, mutually
conjugate ones:
\begin{equation}\label{25}
p=re^{i\theta}, \hspace{5mm} q=re^{-i\theta}.
\end{equation}
Then, instead of $p$ and $q$ we have the parameters $r$ and $\theta$
in terms of which the coefficients of virial expansion (\ref{24})
take the form:
 $$
\widetilde{A}=-\frac{rcos\theta}{2^{5/2}}, \hspace{15mm}
\widetilde{B}=\frac{r^2cos^2\theta}{2^3}-\frac{2r^2(2cos2\theta+1)}{3^{7/2}},
$$\vspace{0mm}
$$
\widetilde{C}=\frac{r^3cos\theta(2cos2\theta+1)}{2^{3/2}3^{3/2}}
-\frac{3r^3(cos3\theta+cos\theta)}{2^6}-\frac{5r^3cos^3\theta}{2^{11/2}},
$$\vspace{0mm}
$$
\widetilde{D}=\frac{7r^4cos^4\theta}{2^6}-\frac{r^4cos^2\theta(2cos2\theta+1)}{2^2\cdot3^{5/2}}+
\frac{r^4cos\theta(cos3\theta+cos\theta)}{2^{7/2}}\,-
$$\vspace{0mm}
$$
\hspace{5mm}-\frac{2r^4(2cos2\theta+1)^2}{3^5}-
\frac{4r^4(2cos4\theta+2cos2\theta+1)}{5^{7/2}}.
$$
Note, at $r=1$ these formulas again reduce to respective results for
the $q$-Bose gas of BM type $q$-bosons, with the phase like
deformation parameter $q=e^{i\theta}$.

Physical meaning of the complex deformation parameters (\ref{25})
can be commented as follows. If one deals with the usual non-ideal
Bose gas, the virial coefficients contain terms responsible for the
(two-particle, three-particle etc.) effective interactions. When
dealing along the same lines with the gas of deformed bosons
(deformed Bose gas), we gain that the substructure of particles or
additional inter-particle interaction is effectively taken into
account. Then the first {\it non-trivial} virial coefficient $A$ or
$\widetilde{A}$ reflects modified two-particle interaction, the
second one $B$ or $\widetilde{B}$ involves modified three-particle
interaction and so on. With two parameters $r$ and $\theta$ at hands
we might hope it is possible to diminish to zero two chosen types of
interaction, say, the two-particle interaction together with the
three-particle one (this would happen if $\widetilde{A}=0$ and
$\widetilde{B}=0$ simultaneously). However, the parameter $r$ (the
modulus) turns out to be of no help for that aim. So, using the
remaining parameter $\theta$ we can find its value(s) for which only
one of the two: $\widetilde{A}$ or $\widetilde{B}$ can be made zero.
If that happens say for $\widetilde{A}$, we may conclude: at this
value of $\theta$ we encounter mutual compensation of the
two-particle interaction present in the usual non-ideal Bose gas
against the additional contribution due to deformation, i.e. due to
the physical reason (finite proper volume or substructure of
particles) effectively taken into account by the deformation.

\begin{figure}[h]
\centerline{\psfig{file=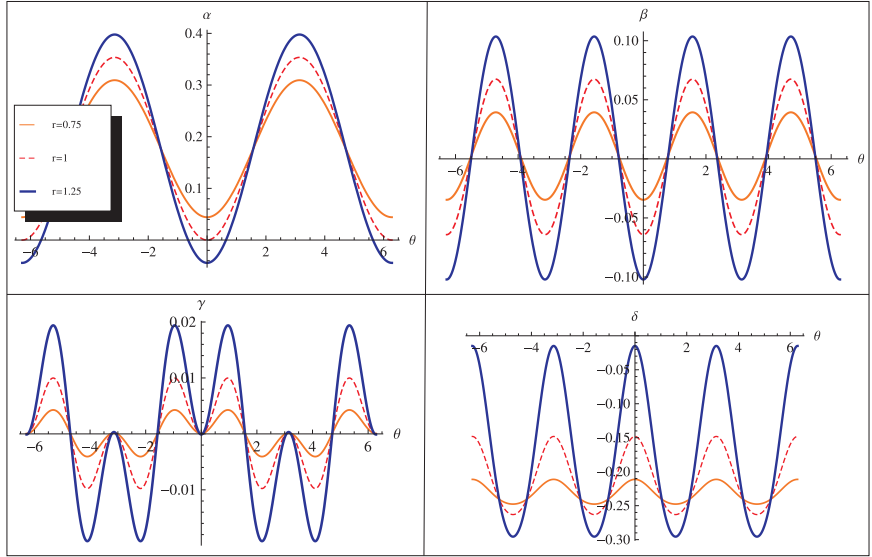,width=11.5cm}} \vspace*{0pt}
\caption{Dependence of $\alpha$, $\beta$, $\gamma$, $\delta$
  on the parameter $\theta$ at $r=0.75, 1, 1.25$.\label{Fig:1}}
\end{figure}

On the other hand, we can find the measure of deviation of "deformed
coefficients" \ from the known virial coefficients \cite{Kubo65},
\cite{Pathria} of the standard Bose gas:
\begin{equation}\label{26}
\alpha=\widetilde{A}-A_0, \hspace{5mm} \beta=\widetilde{B}-B_0,
\hspace{5mm} \gamma=\widetilde{C}-C_0, \hspace{5mm}
\delta=\widetilde{D}-D_0.
\end{equation}
 In Fig. \ref{Fig:1} we plot the dependence of
$\alpha$, $\beta$, $\gamma$, $\delta$ defined in (\ref{26}) on the
parameter $\theta$ at some fixed values of the second parameter $r$.

As seen, for different values of $\theta$ the differences $\alpha$,
$\beta$ and $\gamma$ can be positive, negative or zero, whereas the
difference $\delta$ is always negative (i.e., the net contribution
to the effective five-particle interaction due to deformation is
smaller than the interaction in the non-deformed Bose gas).

\begin{figure}[th]
\centerline{\psfig{file=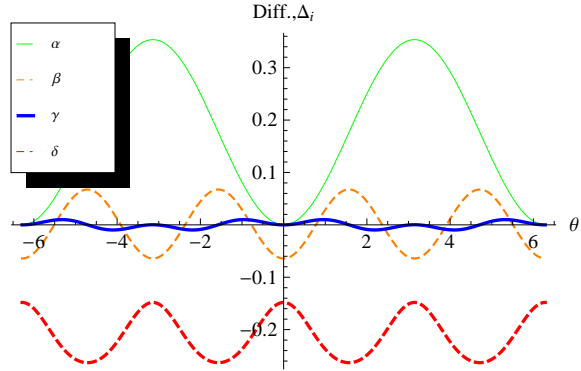,width=8.0cm}} \vspace*{8pt}
\caption{Dependence of the differences $\Delta_i=(\alpha, \beta,
\gamma, \delta)$ from Eq. (\ref{26}) on the parameter $\theta$ at
$r=1$. \label{Fig:2}}
\end{figure}

In Fig. \ref{Fig:2} we plot the dependence of $\alpha$, $\beta$,
$\gamma$, $\delta$ on the parameter $\theta$ at fixed value $r=1$
(that implies the deformed model of BM type $q$-bosons).

\noindent{\it Remark}. From Figs \ref{Fig:1}, \ref{Fig:2} we observe
remarkable thing: for all $\theta\neq0$, we have
$\widetilde{A}>A_0$, but $\widetilde{D}<D_0$. That is the effect of
deformation significantly increases attractive two-particle
interaction for most of values of parameter $\theta$ (at $r\leq 1$
and $\theta\neq 0$). This is in some accord with remark in
\cite{Avancini94} for the case of quon-based deformation of Bose
gas. In the domain of five-particle interaction the picture is
opposite - the deformation modifies the interaction towards
weakening. Although this regime (of high temperatures) is opposite
to that one where the critical $T_c$ is actual (see next Section),
we nevertheless note that the deformation better promotes Bose
condensation that results in increasing $T_c$ (due to the increase
of attractive two-particle interaction). This fact may be of
potential applied value.

\section{\label{sec:4} Bose condensation in $p,q$-Bose gas and
critical temperatures}
 Here we study the opposite situation (to the Subsection 3.1),
 namely, the case of
low temperatures and large densities ($\lambda^3/{\it v}\gg1$).
 Eq. (\ref{21}) gives the equation of
state for $p,q$-deformed Bose gas consisting of $N$ non-relativistic
particles of mass $m$ confined in the volume $V$. To explore the
equation of state in detail we have to find the fugasity $z$ as a
function of temperature, and the specific volume should be found by
solving the equation (\ref{17}) that involves besides
$n_0=\frac{1}{p-q}\ln\Bigl(\frac{1-qz}{1-pz}\Bigr)$, also
$\it{v}\!\!=\!\!V/N$ and the thermal wavelength
$\lambda\!\!=\!\!\sqrt{2\pi\hbar^2/mT}$. Rewrite Eq.~(\ref{17}) as
\begin{equation}\label{27}
\lambda^3\frac{n_0}{V}=\frac{\lambda^3}{{\it v}}-{\widetilde{g}_{3/2}(z;p,q)}.
\end{equation}
We see that the value $n_0/V$ is positive if the temperature and
specific volume obey the following inequality:
\begin{equation}\label{28}
\frac{\lambda^3}{{\it v}}>\widetilde{g}_{3/2}(1;p,q).
\end{equation}
In other words, the large though finite number of particles occupies
the lowest energy level with $\varepsilon_i=0$ (ground state). That
is, the phenomenon of Bose condensation takes place. For given
specific volume ${\it v}$, the critical temperature can be found
from the equality
\begin{equation}\label{29}
\lambda^3_c={\it v}\widetilde{g}_{3/2}(1;p,q).
\end{equation}
Then, the critical temperature $T_c^{(p,q)}$ of the $p,q$-deformed Bose gas results as
 \begin{equation}\label{30}
T_c^{(p,q)}\!= \!\frac{2\pi\hbar^2/m}{[\it{v}
{\widetilde{g}_{3/2}}(1;p,q)]^{2/3}}.
\end{equation}
Moreover, we obtain the relation between the critical temperature of
the considered $p,q$-deformed Bose gas and the $T_c$ of usual gas of
bosons in the form of the ratio:
\begin{equation}  \label{31}
\frac{T_c^{(p,q)}}{T_c}=\Biggr(\!\frac{2.61}{{\widetilde{g}_{3/2}}(1;p,q)}\!\Biggl)^{\!2/3}.
\end{equation}
Note that in the no-deformation limit $p\rightarrow 1$,
$q\rightarrow 1$, the function $\widetilde{g}_{3/2}(1;p,q)$ in
(\ref{30}) goes over into $g_{3/2}(1)=\zeta(\frac{3}{2})\cong2.61$,
as seen from (\ref{19}) at $z=1$. Due to this, the critical
temperature $T_c^{(p,q)}$ of deformed Bose gas reduces to the
critical temperature $T_c$ of usual, non-deformed Bose gas and thus
the ratio is $\bigl(T_c^{(p,q)}\!/\,T_c\bigr)\!\vert_{p=q=1}=1$.

\begin{figure}[th]
\centerline{\psfig{file=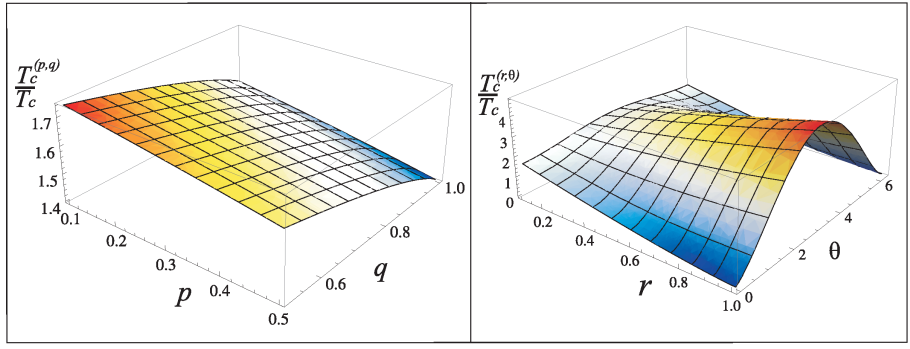,width=12.0cm}}
 \caption{{\it Left}:
The ratio $T_c^{(p,q)}/T_c$ of the critical temperatures given by
Eq. (\ref{31}) as a function of the deformation parameters $p,q$
  such that $0<p\leq1$ and $0<q\leq1$.
{\it Right}: The ratio $T_c^{(r,\theta)}/T_c$ versus the deformation
parameters $r$ and $\theta$, $0<r\leq 1$, $0\leq \theta\leq 2\pi$.
      \label{Fig:3}}
\end{figure}

In Fig. \ref{Fig:3} ({\it left}) we give the plot of (\ref{31}) as a
function of the deformation parameters $p,q$ such that $p\leq1$,
$q\leq1$. This (convex upwards) behavior with respect to $p$ and $q$
differs from that in \cite{Algin2008}. Besides, this ratio increases
with increasing (in the both parameters $p$, $q$) extent of
deformation measured by $1-p$ and $1-q$. In analogy with the above
case of high temperature and low density, where we considered both
real $p$, $q$ and the variant with complex deformation parameters
$p=\overline{q}=r e^{i\theta}$, here in Fig. \ref{Fig:3} ({\it
right}) we also present the picture for
$\bigl(T_c^{(r,\theta)}\!/\,T_c\bigr)$ depending on complex $p$, $q$
through the modulus $r$ and the phase $\theta$.

As seen, the critical temperature $T_c^{(p,q)}$ for the
$p,q$-deformed Bose gas is larger than the critical temperature
$T_c$ for the non-deformed boson gas, at least in the chosen region
$p\leq1$, $q\leq1$ of the deformation parameters (see also Remark
ending Sect. \ref{sec:3}). This fact may play an important role in
future analysis involving real gases. Even more interesting is the
picture for the ratio in Eq. (\ref{31}) (at
$p=\overline{q}=re^{i\theta}$ versus the parameters $r$ and
$\theta$).

\section{\label{sec:5} Deformed Bose gas of Tamm-Dancoff type}
Let us consider thermodynamics of the $q$-Bose gas of Tamm-Dancoff
(TD) type. Recall that the defining (along with (\ref{1}))
commutation relation
\begin{equation}\label{32}
    aa^{\dagger}-qa^{\dagger}a=q^N
\end{equation}
for the TD $q$-bosons, and the $q$-bracket of TD type
\begin{equation}\label{32}
    [X]_{TD}=Xq^{X-1}
\end{equation}
stem from those of the two-parameter $p,q$-deformed model if one
puts $p\!=\!q$, see (\ref{7}). We assume that $0<q\leq1$ in the case
of TD bosons. For some exotic properties of the TD type deformed
oscillator see Ref. \cite{GavrReb07}.

Performing when $\lambda^3/v\ll 1$ same calculations as in Sec.
\ref{sec:3}, in the Tamm-Dancoff case we find the coefficients for
the virial expansion (\ref{24}):

$$ A_{TD}=-\frac{q}{2^{5/2}}=qA_0, \hspace{12mm} C_{TD}=
q^3\biggl(\frac{1}{2^{3/2}3^{1/2}}-\frac{3}{2^{5}}-\frac{5}{2^{11/2}}\biggr)=q^3C_0,
$$
$$
B_{TD}=q^2\biggl(\frac{1}{2^3}-\frac{2}{3^{5/2}}\biggr)=q^2B_0, \hspace{5mm}
D_{TD}=q^4\biggl(\frac{7}{2^{6}}-\frac{1}{2^2\cdot3^{3/2}}+
\frac{1}{2^{5/2}}+\frac{2}{3^{3}}-\frac{4}{5^{5/2}}\biggr)=q^4D_0.
$$
It is worth to note that, due to $q\leq1$, the TD Bose gas virial coefficients are lowered
with respect to non-deformed virial coefficients,
 by the factor
of $q^{k-1}$ in the $k$-th virial coefficient. Accordingly, this weakens in $q^{k-1}$
times the usual $k$-particle effective interaction within the TD $q$-Bose gas.

\begin{figure}[th]
\centerline{\psfig{file=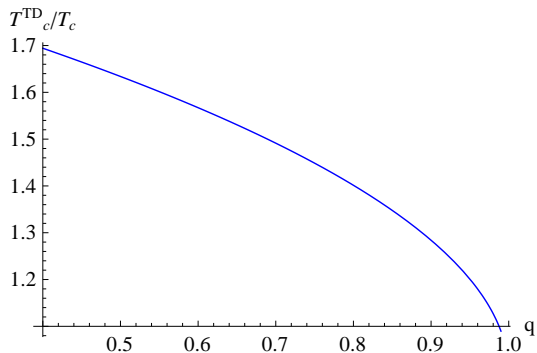,width=7cm}}
 \caption{The ratio
$T_c^{TD}/T_c$
  of the $TD$-type deformed critical temperature $T_c^{TD}$ to the
  non-deformed one $T_c$ versus the deformation parameter
  $q$ where $0<q\leq1$ .\label{Fig:4}}
\end{figure}

In the case of low temperatures and large densities, $\lambda^3/v\gg 1$,
the critical temperature of Bose gas of Tamm-Dancoff type is
\begin{equation}\label{33}
    \frac{T_c^{TD}}{T_c}=\biggl(\frac{2.61}{g_{3/2}^{TD}(1;q)}\biggr)^{2/3},
\end{equation}
where $g_{3/2}^{TD}(1;q)$ is the TD analog of function (\ref{19}) at $z\!=\!1$ and $n\!=\!3/2$.
In general,
\begin{equation}\label{34}
g_n^{TD}(z;q)=\sum_{r=1}^{\infty}\frac{q^{r-1}z^r}{r^{n}}.
\end{equation}
In Fig. \ref{Fig:4}, we plot the ratio (\ref{33}) versus deformation
parameter $q$ for $0<q\leq1$.
As seen, the larger is deformation (deviation $1-q$ from the
non-deformed value $q\!=\!1$), the larger is the critical
temperature $T_c^{TD}$ of deformed bosons of TD type. At $q=1$ the
ratio (\ref{33}) gives 1, as it should be.

\section{\label{sec:6} Concluding remarks}
In this paper we developed some version of the $p,q$-deformed analog
of the Bose gas model ($p,q$-Bose gas model). We studied the
thermodynamics of such a gas at high (low) temperatures and low
(large) densities. To obtain the thermodynamical quantities we
utilized the extension of $q$-calculus based on the direct
$p,q$-generalization of the Jackson derivative.

For high temperatures ($\lambda^3/{\it v}\ll1$), dealing with the
equation of state we have obtained the virial expansion and
correspondingly derived the nontrivial virial coefficients $A$, $B$,
$C$, $D$ all depending on the deformation parameters $p$ and $q$.
Note that these parameters are contained in the virial coefficients
through the $p,q$-numbers only and are thus symmetric under the
exchange $p\leftrightarrow q$. In the limit $p\!=\!q\!=\!1$ the
virial coefficients $A_0$, $B_0$, $C_0$, $D_0$ of usual Bose gas are
recovered. The parameters $p$ and $q$ are taken first to be real,
and then also as complex valued such that
$p\!=\!\overline{q}\!=\!re^{i\theta}$. In these new variables $r$
and $\theta$ we analyzed the virial coefficients as well. The
differences between deformed coefficients and their non-deformed
counterparts have been studied from the viewpoint of their
dependence on the parameter $\theta$. Figures \ref{Fig:1},
\ref{Fig:2} demonstrate: there exist some special value(s) of the
$\theta$-parameter (and $r$) for which any chosen difference from
(\ref{26}) can vanish due to some mutual compensation. On the other
hand, for special fixed value $\theta$ (at certain $r$), the virial
coefficient $\widetilde{A}$, and thus the effective two-particle
interaction can be vanishing. Note that the effective two-particle
interaction (and vanishing of it) means superimposing (and mutual
compensation) of the conventional two-particle interaction, present
for the usual non-deformed non-ideal Bose-gas, and the additional
interaction imported due to the deformation used by us. Evidently,
the same can be said about each from the rest of virial coefficients
(taken alone).

At low temperatures ($\lambda^3/{\it v}\!\gg\!1$) we find both the
critical temperature $T_c^{(p,q)}$ of deformed $p,q$-Bose gas and
the explicit dependence of the ratio $T_c^{(p,q)}\!/\,T_c$ on the
parameters $p$ and $q$ (either taken to be real, or as the pair of
mutually conjugate complex values). Similar results are deduced for
the deformed Bose gas of Tamm-Dancoff type. We have found, for the
whole range of the considered $p$, $q$, that $T_c^{(p,q)}\!>\!T_c$
(such inequality retains in the case of TD-type $q$-Bose gas also).
We hope this feature may have interesting consequences. To conclude,
the issue of Bose condensation in its $p,q$-deformed manifestation
(with $T_c^{(p,q)}$ or $T_c^{(r,\theta)}$ higher than $T_c$) is
reproduced and pictured in Fig. 3, {\it left} and {\it right}. It
would be interesting to make comparison of the obtained results with
the critical $T_c$ of a real gas.

\section*{Acknowledgments}

This research was partially supported by the Special Program of the
Division of Physics and Astronomy of the NAS of Ukraine.

\end{document}